\documentclass[reprint,superscriptaddress,amsmath,amssymb,pra,aps,showpacs]{revtex4-1}

\usepackage{amsmath}
\usepackage {xcolor}
\usepackage{comment}

\usepackage{mathrsfs}
\usepackage{comment}
\usepackage{amssymb}
\usepackage{graphicx}
\usepackage{subfigure}
\usepackage[colorlinks,
            linkcolor=blue,
            anchorcolor=blue,
            citecolor=blue]{hyperref}
\newtheorem{theorem}{Theorem}

\def\ra{\rangle}
\def\la{\langle}

\usepackage{graphicx}
\usepackage{dcolumn}
\usepackage{bm}
\begin{document}


\title{Tightening the entropic uncertainty relations with quantum memory in a multipartite scenario}

\author{Cong Xu}
\email[]{2230501028@cnu.edu.cn}
\affiliation{School of Mathematical Sciences, Capital Normal University, Beijing 100048, China}

\author{Qing-Hua Zhang}
\email[]{qhzhang@csust.edu.cn}
\affiliation{School of Mathematics and Statistics, Changsha University of Science and Technology, Changsha 410114, China}

\author{Tao Li}
\email[]{litao@btbu.edu.cn}
\affiliation{School of Mathematics and Statistics, Beijing Technology and Business University, Beijing 100048, China}

\author{Shao-Ming Fei}
\email[]{feishm@cnu.edu.cn}
\affiliation{School of Mathematical Sciences, Capital Normal University, Beijing 100048, China}

\begin{abstract}

The quantum uncertainty principle stands as a cornerstone and a distinctive feature of quantum mechanics, setting it apart from classical mechanics. We introduce a tripartite quantum-memory-assisted entropic uncertainty relation, and extend the relation to encompass multiple measurements conducted within multipartite systems. The related lower bounds are shown to be tighter than those formulated by Zhang \textit{et al.} [Phys. Rev. A 108, 012211 (2023)]. Additionally, we present generalized quantum-memory-assisted entropic uncertainty relations (QMA-EURs) tailored for arbitrary positive-operator-valued measures (POVMs). Finally, we demonstrate the applications of our results to both the relative entropy of unilateral coherence and the quantum key distribution protocols.

\end{abstract}

\maketitle

\section{Introduction}

The uncertainty relation was initially proposed by Heisenberg \cite{heisenberg1927uber}. Robertson \cite{robertson1929the} derived the well-known product form uncertainty relation for any two observables $M_1$ and $M_2$ with respect to a fixed state $|\psi\rangle$,
\begin{align}\label{eq1}
\Delta M_1 \Delta M_2 \geq \frac{1}{2} |\langle\psi|[M_1, M_2]|\psi\rangle|,
\end{align}
where $\Delta M = \sqrt{\langle\psi|M^2|\psi\rangle - \langle\psi|M|\psi\rangle^2}$ is the standard deviation of an observable $M$ and $[M_1, M_2] = M_1 M_2 - M_2 M_1$.
Here, the lower bound of (\ref{eq1}) is state-dependent and becomes trivial when $|\psi\rangle$ is an eigenvector of either $M_1$ or $M_2$.

To address this issue, Deutsch \cite{deutsch1983uncertainty} introduced the entropic uncertainty relation, which was subsequently improved \cite{kraus1987complementary} and proved by Maassen and Uffink \cite{maassen1988generalized},
\begin{align}\label{eq2}
H(M_1) + H(M_2) \geq q^{M_{1,2}}_{MU},
\end{align}
where $H(M_1) = -\sum\limits_j p_j \log_2 p_j$ denotes the Shannon entropy of the probability distribution $p_j = \langle \varphi_j^1|\rho|\varphi_j^1\rangle$, and $q^{M_{1,2}}_{MU} = -\log_2 c_{\max}^{1,2}$ with $c_{\max}^{1,2} = \max_{j,k} |\langle \varphi_j^1 | \varphi_k^2 \rangle|^2$. Here, $\{|\varphi_j^1\rangle\}$ and $\{|\varphi_k^2\rangle\}$ represent the eigenvectors of observables $M_1$ and $M_2$, respectively. This uncertainty relation inspired the initial proposal for quantum cryptography \cite{10.1145/1008908.1008920}. However, this uncertainty relation was proven to be not sufficient in proving the security of encryption, as it does not account for the possible entanglement \cite{RevModPhys.81.865} between the eavesdropper and the system being measured. Berta \textit{et al.} \cite{berta2010uncertainty} addressed this gap by generalizing the uncertainty relation (\ref{eq2}) to the case involving quantum memory. For any bipartite state $\rho_{AB}$, Bob's uncertainty about the measurement outcomes is bounded by the following inequality,
\begin{align}\label{eq3}
S(M_1 | B) + S(M_2 | B) \geq q^{M_{1,2}}_{MU} + S(A | B),
\end{align}
where $S(M_1 | B) = S(\rho_{M_1 B}) - S(\rho_B)$ (similarly for $S(M_2|B)$) denotes the conditional entropy with $\rho_{M_1 B}=\sum_{j}(|\varphi_j^1\rangle\langle \varphi_j^1|\otimes \mathrm{I_{B}})\rho_{AB}(|\varphi_j^1\rangle\langle \varphi_j^1|\otimes \mathrm{I_{B}})$, quantifying the uncertainty about the outcome of measurement $M_1$ given the information stored in a quantum memory $B$.
$S(\rho)=-\mathrm{tr} \rho\log\rho$ denotes the von Neumann entropy. $S(A|B)=S(\rho_{AB})-S(\rho_{B})$ is related to the entanglement between the measured particle $A$ and the quantum memory $B$.

Renes\textit{et al.} \cite{renes2009conjectured} considered tripartite quantum-memory-assisted entropic uncertainty relation (QMA-EUR), and proposed a relation for a pair of conjugate observables related by a Fourier transform. Berta \textit{et al.} \cite{berta2010uncertainty} generalized this scenario to two arbitrary measurements $M_1$ and $M_2$,
\begin{equation}\label{eq4}
S(M_1 | B) + S(M_2 | C) \geq q^{M_{1,2}}_{MU},
\end{equation}
where the three players, Alice, Bob and Charlie, share a tripartite state $\rho_{ABC}$, and the particles $A$, $B$ and $C$ are distributed to Alice, Bob and Charlie, respectively. Alice randomly selects one of the two measurements $M_1$ and $M_2$, and informs Bob and Charlie of her choice $M \in \{M_1, M_2\}$. If Alice chooses $M_1$, Bob's objective is to minimize his uncertainty about $M_1$. If she chooses $M_2$, Charlie aims to minimize his uncertainty about $M_2$.

The universality of inequality (\ref{eq3}) has catalyzed a range of applications, encompassing entanglement witnesses \cite{berta2010uncertainty, prevedel2011experimental, li2011experimental, berta2014entanglement, huang2010entanglement, hu2012quantum}, EPR steering \cite{sun2018demonstration, schneeloch2013einstein}, quantum metrology \cite{giovannetti2011advances}, quantum key distribution \cite{koashi2009simple, berta2010uncertainty, PhysRevLett.106.110506, tomamichel2012tight}, quantum cryptography \cite{dupuis2014entanglement, konig2012unconditional} and quantum randomness \cite{vallone2014quantum}. Recently, the lower bounds of QMA-EUR, as encapsulated by inequalities (\ref{eq3}) and (\ref{eq4}), have been further refined and generalized \cite{coles2014improved, pati2012quantum, adabi2016tightening, ming2020improved, wu2022tighter}.

The exploration of QMA-EUR for multiple measurements in multipartite has garnered significant interest \cite{liu2015entropic, zhang2015entropic, coles2017entropic, hu2013competition, chen2018improved, xie2021optimized, xiao2016strong, wu2022tighter}. In 2015, Liu \textit{et al.} \cite{liu2015entropic} extended the ``uncertainty game" to multiple measurement contexts within bipartite systems. By reordering the measurements, Zhang \textit{et al.} \cite{zhang2015entropic} derived a tighter lower bound compared to the theorem 2 in \cite{liu2015entropic}. Leveraging majorization theory and symmetric group actions, Xiao \textit{et al.} \cite{xiao2016strong} introduced an admixture lower bound that surpasses Zhang \textit{et al.}'s results. Dolatkhah \textit{et al.} \cite{dolatkhah2019tightening} subsequently enhanced the lower bounds proposed by Zhang \textit{et al.}, Liu \textit{et al.} and Xiao \textit{et al.}, transitioning the entropic uncertainty relation from the absence to the presence of quantum memory. Taking into account the additivity of (\ref{eq3}), Xie \textit{et al.} \cite{xie2021optimized} developed a simply constructed bound (SCB), which improved upon Liu \textit{et al.}'s results for any multiple mutually unbiased basis (MUB) measurements \cite{doi:10.1142/S0219749910006502, PhysRevLett.122.050402}. Wu \textit{et al.} \cite{wu2022tighter} generalized the entropic uncertainty relation to multiple measurements across multipartite systems. Zhang \textit{et al.} \cite{PhysRevA.108.012211} further refined Wu \textit{et al.}'s results and extended the QMA-EUR to encompass $m$ measurements and $n$ memories ($n \leq m$).

The paper is structured as follows. In Section 2, we introduce a tripartite QMA-EUR and generalize it to arbitrary multi-measurements in multipartite systems. Section 3 presents the generalized QMA-EURs for arbitrary positive-operator-valued measures (POVMs). The application of our results to quantum coherence and quantum secret key rate is discussed in Section 4. Finally, we summarize in Section 5.

\section{Tripartite QMA-EUR and generalized QMA-EURs for multiple measurements in multipartite systems}

We introduce a tripartite QMA-EUR and extend it to the scenario involving \( m \) measurements and \( n \) memories (\( n \leq m \)).
Consider two measurements $M_1$ and $M_2$ given by eigenvectors $\{|\varphi_j^1\rangle\}$ and $\{|\varphi_k^2\rangle\}$, respectively, in a quantum system $A$ with finite dimension $d$.
 Coles \textit{et al.} \cite{coles2014improved} enhanced the lower bounds of inequalities (\ref{eq3}) and (\ref{eq4}) as follows,
\begin{equation}\label{eq5}
S(M_1|B) + S(M_2|B) \geq Q^{M_{1,2}} + S(A|B),
\end{equation}
\begin{equation}\label{eq6}
S(M_1|B) + S(M_2|C) \geq Q^{M_{1,2}},
\end{equation}
where \( Q^{M_{1,2}} \) is one of \( \widetilde{q}^{M_{1,2}} \), \( q^{M_{1,2}}(\rho_A) \) and \( q^{M_{1,2}} \).
Specifically,
\begin{equation}\label{eq7}
\widetilde{q}^{M_{1,2}} = q^{M_{1,2}}_{MU} + \frac{1}{2}(1 - \sqrt{c^{1,2}_{\max}})\log_2\frac{c^{1,2}_{\max}}{c^{1,2}_2},
\end{equation}
where \( c^{1,2}_2 \) represents the second largest element of the matrix \( [c_{j,k}^{1,2}] \).
\begin{equation}\label{eq8}
q^{M_{1,2}}(\rho_A) = \max\{q(\rho_{A}, M_1, M_2), q(\rho_{A}, M_2, M_1)\},
\end{equation}
with \( q(\rho_{A}, M_1, M_2) = \sum\limits_j p^{M_1}_j \log_2(1 / \max\limits_k c^{1,2}_{jk}) \) and \( q(\rho_{A}, M_2, M_1) = \sum\limits_k p^{M_2}_k \log_2(1 / \max\limits_j c^{1,2}_{jk}) \), where $p^{M_1}_j=\langle \varphi_j^1|\rho_{A}|\varphi_j^1\rangle$, $p^{M_2}_k=\langle \varphi_k^2|\rho_{A}|\varphi_k^2\rangle$.
\( q^{M_{1,2}} \) is defined as the minimum value taken over all possible states \( \rho_A \) \cite{coles2014improved},
\begin{equation}\label{eq9}
q^{M_{1,2}} = \max_{0 \leq p \leq 1} \lambda_{\min}[\Delta(p)],
\end{equation}
where \( \lambda_{\min}[\cdot] \) denotes the smallest eigenvalue, and \( \Delta(p) = p \Delta_{M_1M_2} + (1 - p) \Delta_{M_2M_1} \). Here, \( \Delta_{M_1M_2} = \sum\limits_j \log_2(1 / \max\limits_k c^{1,2}_{jk}) |\varphi_j^1\rangle \langle \varphi_j^1| \) and \( \Delta_{M_2M_1} = \sum\limits_k \log_2(1 / \max\limits_j c^{1,2}_{jk}) |\varphi_k^2\rangle \langle \varphi_k^2| \). Note that \( Q^{M_{1,2}} \geq q^{M_{1,2}}_{MU} =\log_2 (1/c_{\max}^{1,2})\) for almost all choices of the eigenvectors of $M_i$ and $M_j$. For \( Q^{M_{1,2}} \) is one of \( \widetilde{q}^{M_{1,2}} \) and  \( q^{M_{1,2}}(\rho_A) \) the inequality becomes an equality when the system $A$ is a qubit ($d = 2$), and if we take $Q^{M_{1,2}}= q^{M_{1,2}}$, the equation holds when the function $\lambda_{\min}[\Delta(p)]$ is independent of $p$ \cite{coles2014improved}.

In the absence of the quantum memory \( B \), the inequality (\ref{eq5}) simplifies to,
$H(M_1) + H(M_2) \geq Q^{M_{1,2}} + S(A)$. By leveraging the relation between conditional entropy and Holevo quantity \cite{Nielsen_Chuang_2010},
$S(M_i|B) = H(M_i) - \mathcal{I}(M_i:B)$, where $\mathcal{I}(M_i:B)=H(M_i) + S(\rho_B) - S(\rho_{M_i B})$ is the Holevo quantity, we have the following Theorem.

\begin{theorem}\label{qmaeurthm1} The following tripartite QMA-EUR inequality holds,
\begin{equation}\label{eq10}
\begin{aligned}
S(M_1|B) + S(M_2|C) \geq Q^{M_{1,2}} + \max\{0, \Delta\},
\end{aligned}
\end{equation}
where \(\Delta = S(A) - \mathcal{I}(M_1:B) - \mathcal{I}(M_2:C)\).
\end{theorem}

\noindent \textit{Proof.} Since
\begin{align}\label{eq11}
&S(M_1|B) + S(M_2|C) \\ \notag
=& H(M_1) + H(M_2) - \mathcal{I}(M_1:B) -\mathcal{I}(M_2:C)\\ \notag
\geq& Q^{M_{1,2}} + S(A) - \mathcal{I}(M_1:B) - \mathcal{I}(M_2:C),
\end{align}
combining inequalities (\ref{eq11}) and (\ref{eq6}), we prove the inequality (\ref{eq10}) through optimization over the two bounds. \(\Box\)

Now consider a multipartite state \(\rho_{AB_1B_2\ldots B_n}\) associated to Alice, Bob$_1$, Bob$_2$, \(\ldots\), Bob$_n$, respectively. Employing the scheme proposed in \cite{PhysRevA.108.012211}, all participants agree on an \(m\)-tuple of measurements \(\mathbf{M} = \{M_i\}_{i=1}^m\). Define \(n\) (with \(n \leq m\)) non-empty subsets \(\mathbf{S}_t\) of \(\mathbf{M}\), such that \(\bigcup_{t=1}^n \mathbf{S}_t = \mathbf{M}\) and \(\mathbf{S}_s \cap \mathbf{S}_t = \emptyset\) for \(s \neq t\), with each \(\mathbf{S}_t\) corresponding to Bob$_t$. Alice randomly performs one of the measurements in \(\mathbf{S}_t\) and informs Bob$_t$ of her choice. In particular, if Alice's measurement outcomes can be perfectly predicted by Bob$_t$, Bob$_t$ wins the game. Zhang \textit{et al.} \cite{PhysRevA.108.012211} introduced the following multipartite QMA-EUR,
\begin{equation}\label{eq12}
\begin{aligned}
\sum_{t=1}^n \sum_{M_i \in \mathbf{S}_t} S(M_i|B_t) \geq & -\frac{1}{m-1} \log_2\left(\prod_{i<j}^m c_{i,j}\right)\\ +  \sum_t \frac{m_t(m_t-1)}{2(m-1)} S(A|B_t)
&+ \max\{0, \delta_{mn}\},
\end{aligned}
\end{equation}
where
\begin{equation*}
\begin{aligned}
\delta_{mn} = & \frac{m(m-1) - \sum_{t=1}\limits^n m_t(m_t-1)}{2(m-1)} S(A)\\
 +& \sum_{t=1}^n \frac{m_t(m_t-1)}{2(m-1)} \mathcal{I}(A:B_t)\\
-& \sum_{t=1}^n \sum_{M_i \in \mathbf{S}_t} \mathcal{I}(M_i:B_t)
\end{aligned}
\end{equation*}
and
\begin{equation}\label{eq13}
\begin{aligned}
\sum_{t=1}^n \sum_{M_i \in \mathbf{S}_t} S(M_i|B_t) \geq &- \frac{1}{m-1} \log_2\left(\prod_{i<j}^m c_{i,j}\right)\\ +  \sum_t \frac{m_t(m_t-1)}{2(m-1)} S(A|B_t)&+\max\{0, \delta_{mn}^\prime\}
\end{aligned}
\end{equation}
with
\begin{equation*}
\begin{aligned}
\delta_{mn}^\prime = & \frac{1}{m-1} \log_2\frac{\left(\prod\limits_{i<j}^m c_{i,j}\right)}{b^{m-1}} + (m-1) S(A)\\
-& \sum_{t=1}^n \frac{m_t(m_t-1)}{2(m-1)} S(A) + \sum_{t=1}^n \frac{m_t(m_t-1)}{2(m-1)} \mathcal{I}(A:B_t)\\
-& \sum_{t=1}^n \sum_{M_i \in \mathbf{S}_t} \mathcal{I}(M_i:B_t),
\end{aligned}
\end{equation*}
\(M_i\) denotes the \(i\)-th measurement on subsystem \(A\) and \(B_t\) represents the \(t\)-th quantum memory in the multipartite system, and \(c_{i,j} = \max\limits_{k,l} |\langle \varphi_k^i | \varphi_l^j \rangle|^2\) with eigenvectors \(\{|\varphi_k^i \rangle\}\) and \(\{|\varphi_l^j \rangle\}\) of \(M_i\) and \(M_j\), respectively, $b=\max \limits_{k_m}\{\sum\limits_{k_2 \sim k_{m-1}} \max \limits_{k_1} |\langle \varphi_{k_1}^1 | \varphi_{k_2}^2\rangle |^2 \prod_{i=2}^{m-1} |\langle \varphi_{k_i}^i| \varphi_{k_{i+1}}^{i+1}\rangle |^2 \}$, \(\mathcal{I}(A:B_t) = S(\rho_A) + S(\rho_{B_t}) - S(\rho_{A B_t})\) is the mutual information, and \(\mathcal{I}(M_i:B_t) = H(M_i) + S(\rho_{B_t}) - S(\rho_{M_i B_t})\) is the Holevo quantity.

Following the approach in \cite{PhysRevA.108.012211, wehner2010entropic, schwonnek2018additivity}, we derive two tighter lower bounds for QMA-EURs concerning \(m\) measurements within the framework of \(n\) memories.
\begin{theorem}\label{theorem2}
By incorporating mutual information and Holevo quantities, the optimized QMA-EUR for \(m\) measurements within the context of \(n\) memories can be expressed as
\begin{equation}\label{eq14}
\begin{aligned}
\sum_{t=1}^n \sum_{M_i \in \mathbf{S}_t} S(M_i|B_t) \geq & \frac{1}{m-1} \sum_{i<j} Q^{M_{i,j}}\\
 +& \frac{1}{m-1} \sum_{t=1}^n \frac{m_t(m_t-1)}{2} S(A|B_t) \\
 +& \max\{0, \delta_{mn}\},
\end{aligned}
\end{equation}
where
\(\delta_{mn}\) is defined in equation (\ref{eq12}).
\end{theorem}

\noindent {\textit{Proof.}} Employing the identity
\(S(M_i|B_s) = H(M_i) - \mathcal{I}(M_i:B_s)\)
and
\(S(M_j|B_t) = H(M_j) - \mathcal{I}(M_j:B_t)\),
we derive
\begin{align*}
&S(M_i|B_s) + S(M_j|B_t)\\ \notag
&= H(M_i) - \mathcal{I}(M_i:B_s) + H(M_j) - \mathcal{I}(M_j:B_t)\\
&\geq Q^{M_{i,j}} + S(A) - \mathcal{I}(M_i:B_s) - \mathcal{I}(M_j:B_t).
\end{align*}
For all \(i < j\), we obtain \(\displaystyle \frac{m(m-1)}{2}\) similar inequalities. Summarizing these inequalities and dividing both sides by \(m-1\), we get
\begin{equation}\label{eq15}
\begin{aligned}
\sum_{t=1}^n \sum_{M_i \in \mathbf{S}_t} S(M_i|B_t) \geq & \frac{1}{m-1} \sum_{i<j} Q^{M_{i,j}} + \frac{m}{2} S(A) \\
-& \sum_{t=1}^n \sum_{M_i \in \mathbf{S}_t} \mathcal{I}(M_i:B_t).
\end{aligned}
\end{equation}

The QMA-EURs (\ref{eq5}) and (\ref{eq6}) can be reformulated as:
\begin{equation*}
S(M_i|B_t) + S(M_j|B_s) \geq \left\{
\begin{aligned}
& Q^{M_{i,j}} + S(A|B_t), & \text{if } t = s, \\
& Q^{M_{i,j}}, & \text{if } t \neq s.
\end{aligned}
\right.
\end{equation*}
By summing these inequalities for all \(i < j\) and subsequently dividing both sides of the summed inequality by \(m-1\), we obtain
\begin{equation}\label{eq16}
\begin{aligned}
\sum_{t=1}^n \sum_{M_i \in \mathbf{S}_t} S(M_i|B_t) \geq  &\frac{1}{m-1} \sum_{i<j} Q^{M_{i,j}}\\
+& \frac{1}{m-1} \sum_{t=1}^n \frac{m_t(m_t-1)}{2} S(A|B_t),\\
\end{aligned}
\end{equation}
where \(m_t\) denotes the cardinality of \(\mathbf{S}_t\).

Combining these inequalities (\ref{eq15}) and (\ref{eq16}), we complete the proof. \(\Box\)

Let $\rho$ be a mixed state in a $d$ dimensional Hilbert space, and $M_i$ $(i=1,2,\ldots,m)$ be $m$ measurements with a set of orthonormal eigenvectors $\{|\varphi^i_{i_m}\rangle\}$ $(i_m=1,2,\ldots,d)$.
Denote  $p^i_{i_m}=\langle \varphi^i_{i_m}|\rho|\varphi^i_{i_m}\rangle$ the probability distributions obtained by measuring $\rho$ with respect to bases $\{|\varphi^i_{i_m}\rangle\}$.
By employing majorization theory and actions of the symmetric group, Xiao \textit{\textit{et al.}}~\cite{xiao2016strong} presented an admixture bound of the entropic uncertainty relation for multiple measurements:
\begin{equation}\label{eq19}
\sum_{i=1}^m H(M_i) \geq -\frac{1}{m} \omega \beta + (m-1) S(A),
\end{equation}
where $\beta$ is the $d^m$ dimension vector  $(\log(\omega\cdot \mathfrak{U}_{i_1,i_2\cdots i_m}))^\downarrow$ defined as follows. The symbol $\downarrow$ stands for arranging the components in descending order. For each multi-index $(i_1,i_2\cdots,i_m)$, the $d^m$ dimensional vector $\mathfrak{U}_{i_1,i_2\cdots i_m}$ has entries of the form $c(1,2,\cdots m)c(2,3, \cdots,1)\cdots c(m,1,\cdots,m-1)$ sorted in decreasing order with
respect to the indices $(i_1,i_2,\cdots,i_m)$, where $c(1,2,\cdots m)=\sum_{i_2,\cdots,i_{m-1}} \max \limits_{i_1} c(\varphi_{i_1}^1,\varphi_{i_2}^2)\cdots c(\varphi_{i_{m-1}}^{m-1},\varphi_{i_m}^m)$. The majorization vector bound $\omega$  for probability tensor distributions $(p^1_{i_1}p^2_{i_2}\cdots p^m_{i_m})_{i_1,i_2,\cdots,i_m}$ of a state $\rho$ is the $d^m$-dimensional vector $\omega=(\Omega_1,\Omega_2-\Omega_1,\cdots,1-\Omega_a,0,\cdots,0)$, where
\begin{equation*}
\Omega_k=(\frac{s_k}{m})^m.
\end{equation*}
We have $\Omega_1\leq\Omega_2\cdots\Omega_a<1$ for some integer $a\leq d^m-1$ with $\Omega_{a+1}=1$. The more detailed definition of $s_k$, see \cite{xiao2016strong}.

\begin{theorem}\label{qmaeurthm3}
For the case of \(m\) measurements involving \(n\) memories, the following entropic uncertainty relation holds,
\begin{equation}\label{eq17}
\begin{aligned}
\sum_{t=1}^n\sum_{M_i\in \mathbf{S}_t} S(M_i|B_t)\geq &\frac{1}{m-1} \sum_{i<j} Q^{M_{i,j}}\\
+&\frac{1}{m-1}\sum_t\frac{m_t(m_t-1)}{2}S(A|B_t)\\
+&\max\{0,\delta_{mn}^{''}\},
\end{aligned}
\end{equation}
where
\begin{equation*}
\begin{aligned}
\delta_{mn}^{''}=&-\frac{1}{m}\omega\beta+(m-1)S(A)-\sum_{t=1}^n\frac{m_t(m_t-1)}{2(m-1)}S(A)\\
&+\sum_{t=1}^n\frac{m_t(m_t-1)}{2(m-1)}\mathcal{I}(A:B_t)\\
&-\frac{1}{m-1} \sum_{i<j}Q^{M_{i,j}}-\sum_{t=1}^n\sum_{M_i\in \mathbf{S}_t} \mathcal{I}(M_i:B_t).
\end{aligned}
\end{equation*}
\end{theorem}

\noindent \textit{Proof.} From the identity \(S(M_i|B_t) = H(M_i) - \mathcal{I}(M_i:B_t)\), we have
\begin{equation}\label{eq18}
\begin{aligned}
\sum_{t=1}^n \sum_{M_i \in \mathbf{S}_t} S(M_i|B_t) = &\sum_{i=1}^m H(M_i)
-\sum_{t=1}^n \sum_{M_i \in \mathbf{S}_t} \mathcal{I}(M_i:B_t).
\end{aligned}
\end{equation}

 Integrating the inequality~(\ref{eq19}) with equation~(\ref{eq18}) we derive a more stringent lower bound,
\begin{equation}\label{eq20}
\begin{aligned}
\sum_{t=1}^n \sum_{M_i \in \mathbf{S}_t} S(M_i|B_t) \geq &  -\frac{1}{m} \omega \beta + (m-1) S(A) \\
&- \sum_{t=1}^n \sum_{M_i \in \mathbf{S}_t} \mathcal{I}(M_i:B_t).
\end{aligned}
\end{equation}
Combining the inequalities (\ref{eq20}) and (\ref{eq16}), we finish the proof. \(\Box\)

It is interesting to compare our Theorems 2 and 3 with the results given by Zhang \textit{\textit{et al.}}~\cite{PhysRevA.108.012211} and compare the lower bounds of Theorems 2 and 3.
For convenience, we denote the right-hand sides of the inequalities (\ref{eq12}), (\ref{eq13}), (\ref{eq14}) and (\ref{eq17}) by \(lb1\), \(lb2\), \(LB1\) and \(LB2\), respectively. Through straightforward calculation, we obtain that
\begin{equation*}
\begin{aligned}
LB1 - LB2
=& \max\{0, \delta_{mn}\} - \max\{0, \delta_{mn}^{''}\}\\
=&\frac{2-m}{2}S(A)+\frac{1}{m}\omega\beta+\frac{1}{m-1}\sum_{i<j}Q^{M_{i,j}}.
\end{aligned}
\end{equation*}
\begin{equation*}
\begin{aligned}
LB1 - lb1 = &\frac{1}{m-1} \sum_{i<j} Q^{M_{i,j}} + \frac{1}{m-1} \log_2 \left( \prod_{i<j}^m c_{i,j} \right), \\
\end{aligned}
\end{equation*}
\begin{equation*}
\begin{aligned}
LB1 - lb2 = &\frac{1}{m-1} \sum_{i<j} Q^{M_{i,j}} + \frac{1}{m-1} \log_2 \left( \prod_{i<j}^m c_{i,j} \right)+\\
& \max\{0, \delta_{mn}\} - \max\{0, \delta_{mn}'\} \\
=& \frac{2-m}{2}S(A)+\frac{1}{m-1}\sum_{i<j} Q^{M_{i,j}}+\log_2 b,
\end{aligned}
\end{equation*}
\begin{align*}
LB2-lb1
=& \frac{1}{m-1} \sum_{i<j} Q^{M_{i,j}} + \frac{1}{m-1} \log_2 \left( \prod_{i<j}^m c_{i,j} \right)\\
 +& \max\{0, \delta_{mn}^{''}\} - \max\{0, \delta_{mn}\}\\
=& \frac{m-2}{2}S(A)+\frac{1}{m-1} \log_2 \left( \prod_{i<j}^m c_{i,j} \right)-\frac{1}{m}\omega\beta
\end{align*}
and
\begin{align*}
LB2 - lb2
=& \frac{1}{m-1} \sum_{i<j} Q^{M_{i,j}} + \frac{1}{m-1} \log_2 \left( \prod_{i<j}^m c_{i,j} \right)\\
 +& \max\{0, \delta_{mn}^{''}\} - \max\{0, \delta_{mn}'\}\\
=& -\frac{1}{m}\omega\beta+\log_{2}b,
\end{align*}
where $$b=\max \limits_{k_m}\{\sum\limits_{k_2 \sim k_{m-1}} \max \limits_{k_1} |\langle \varphi_{k_1}^1 | \varphi_{k_2}^2\rangle |^2 \prod_{i=2}^{m-1} |\langle \varphi_{k_i}^i| \varphi_{k_{i+1}}^{i+1}\rangle |^2 \}$$ is the same as in inequality (\ref{eq13}), and the last equalities in the above equations hold when $\delta_{mn},\delta_{mn}^{''}, \delta_{mn}'\geq0$. In \cite{xiao2016strong}, Xiao \textit{et al.} presented an admixture lower bound that surpassed the result achieved by Liu \textit{et al.}'s~\cite{liu2015entropic}, which ensures that $LB2-lb2$ is non-negative. Combining with $Q^{M_{i,j}} \geq q^{M_{i,j}}_{MU}$, we conclude that our lower bounds $LB1$ and $LB2$ are tighter than $lb1$ and $lb2$, respectively.

Two groups of orthonormal bases $\{|\varphi^i_{i_m}\rangle\}$ and $\{|\varphi^j_{i_n}\rangle\}$ in a $d$-dimension Hilbert space are
called mutually unbiased if  $|\langle \varphi^i_{i_m}|\varphi^j_{i_n}\rangle|^2=1/d$ holds for all $i_m$ and $i_n$\cite{doi:10.1142/S0219749910006502, PhysRevLett.122.050402}. In fact, when we consider mutually unbiased measurements (MUMs), we obtain $Q^{M_{i,j}}=q^{M_{i,j}}_{MU}$ and $\omega\beta=-m\log_2d$, i.e., $LB1=lb1$, $LB2=lb2$. We also find that
\begin{equation*}
\begin{aligned}
lb1 - lb2
=&\delta_{mn}'-\delta_{mn}\\
=&\frac{m-2}{2}(\log_2d-S(A))\geq0,
\end{aligned}
\end{equation*}
\begin{equation*}
\begin{aligned}
LB1 - LB2
=&\frac{2-m}{2}S(A)+\frac{1}{m}\omega\beta+\frac{1}{m-1}\sum_{i<j}Q^{M_{i,j}}\\
=&\frac{2-m}{2}S(A)+\frac{1}{m}\omega\beta+\frac{m}{2}\log_2d\\
=&\frac{m-2}{2}(\log_2d-S(A))\geq0.
\end{aligned}
\end{equation*}
The above inequalities are saturated when $m=2$ or $\log_2d=S(A)$ and are strictly larger for $m\geq3$.


To further illustrate the effectiveness of our results, we present the following examples and consider the values of $LB1-lb1$, $LB1-lb2$, $LB2-lb1$, $LB2-lb2$ and $LB1-LB2$. For convenience, we take $Q^{M_{i,j}}=\widetilde{q}^{M_{1,2}}$ in the examples of this paper.

 \textbf{\textit{Example 1}}
Consider the three-qutrit state $|\Phi\rangle=\displaystyle\frac{1}{\sqrt{3}}(|000\rangle+|111\rangle+|222\rangle)$ and three measurements $M_1$, $M_2$ and $M_3$ in a three-dimensional Hilbert space with the following eigenvectors,
$|\varphi^1_1 \rangle=(\sqrt{a},\sqrt{1-a},0)^T$, $|\varphi^1_2 \rangle=(e^{\mathrm{i}\phi}\sqrt{1-a}, -e^{\mathrm{i}\phi}\sqrt{a},0)^T$, $|\varphi^1_3 \rangle=(0,0,1)^T$, $|\varphi^2_1 \rangle=\frac{1}{\sqrt3}(1,1,1)^T$, $|\varphi^2_2 \rangle=\frac{1}{\sqrt3}(1,\omega^\ast,\omega)^T$, $|\varphi^2_3 \rangle=\frac{1}{\sqrt3}(1,\omega,\omega^\ast)^T$,
$|\varphi^3_1 \rangle=\frac{1}{\sqrt3}(1,1,\omega^\ast)^T$, $|\varphi^3_2 \rangle=\frac{1}{\sqrt3}(1,\omega,\omega)^T$, $|\varphi^3_3 \rangle=\frac{1}{\sqrt3}(1,\omega^\ast,1)^T$, where $\omega=e^{\frac{2}{3}\pi\mathrm{i}}$, $\omega^\ast$ represents the conjugate of $\omega$. For the convenience of calculation, we take $a=0.95$ , $\phi=\pi$ and obtain $LB1-LB2=0.0736$, $LB1-lb1=0.0366$, $LB1-lb2=0.4366$, $LB2-lb1=-0.0370$,
$LB2-lb2=0.3630$. It can be seen that for this example, $LB1>lb1>LB2>lb2$, i.e., our lower bound $LB1$ is tighter than the previous results $lb1$ and $lb2$ and $LB1>LB2$.


\textbf{\textit{Example 2}}
Consider the three-qutrit $|\Phi\rangle=\displaystyle\frac{1}{\sqrt{3}}(|012\rangle+|120\rangle+|201\rangle)$ and three measurements $M_1$, $M_1$ and $M_3$ in a three-dimensional Hilbert space with the following eigenvectors, $|\varphi^1_1 \rangle=(1,0,0)^T$, $|\varphi^1_2 \rangle=(0,0,1)^T$, $|\varphi^1_3 \rangle=(0,-1,0)^T$,
$|\varphi^2_1 \rangle=(-\frac{1}{3},-\frac{2}{3},\frac{2}{3})^T$, $|\varphi^2_2 \rangle=(-\frac{2}{\sqrt{5}},\frac{1}{\sqrt{5}},0)^T$, $|\varphi^2_3 \rangle=(\frac{2}{3\sqrt{5}},\frac{4}{3\sqrt{5}},\frac{5}{3\sqrt{5}})^T$,
$|\varphi^3_1 \rangle=(\sqrt{a},\sqrt{1-a},0)^T$, $|\varphi^3_2 \rangle=(e^{\mathrm{i}\phi}\sqrt{1-a}, -e^{\mathrm{i}\phi}\sqrt{a},0)^T$, $|\varphi^3_3 \rangle=(0,0,1)^T$. For the convenience of calculation, we take $\phi=\pi$, $1\geq a\geq0$ and obtain the following Figures.
\begin{figure}[tp]
  \centering
  \includegraphics[width=8 cm]{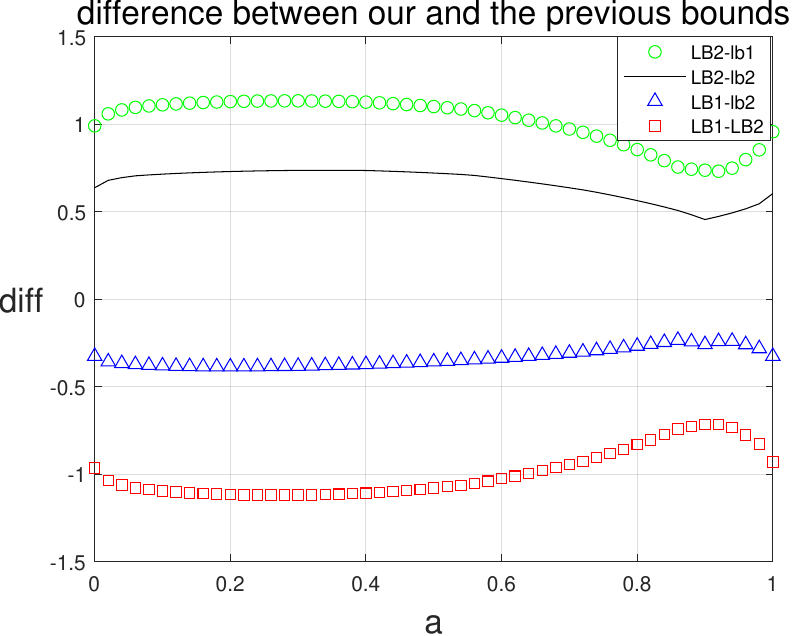}\\
\caption{
The horizontal axis represents the parameter $a$, while the vertical axis represents the difference (diff) between our results and the previous results. The green (circle), black (solid), blue (triangle) and the red (square) curves represent the values of $LB2-lb1$, $LB2-lb2$, $LB1-lb2$ and $LB1-LB2$, respectively.    \label{fig:Fig1}}
\end{figure}
\begin{figure}[tp]
  \centering
  \includegraphics[width=8 cm]{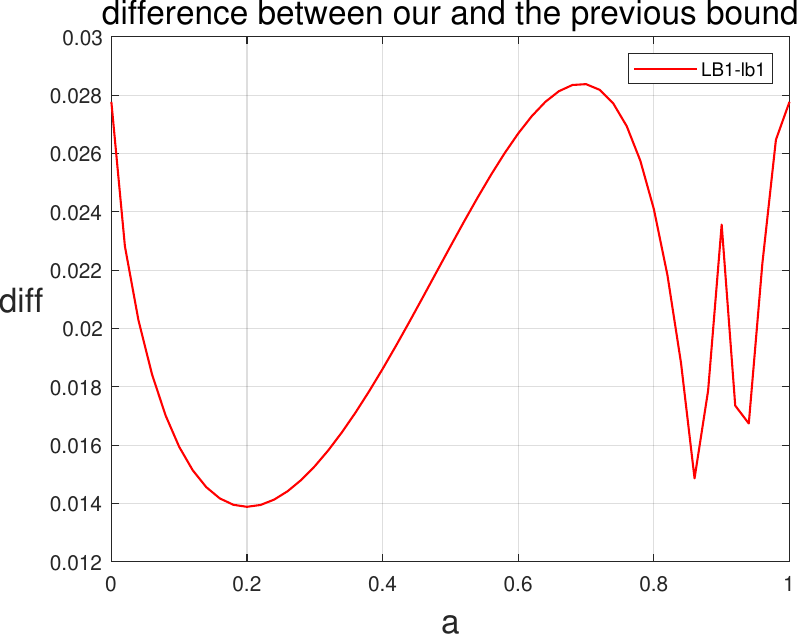}\\
\caption{
The horizontal axis represents the parameter $a$, while the vertical axis represents the difference (diff) between our result and the previous result. This difference (red solid curve) remains above the horizontal axis, indicating that our result is consistently better than the previous one.\label{fig:Fig2}}
\end{figure}



It can be seen from the Figure $1$ and Figure $2$, $LB2>lb1$, $LB2>lb2$, $lb2>LB1$, $LB2>LB1$, that is, $LB2>lb2>LB1>lb1$. It means that our lower bound $LB2$ is tighter than the previous results $lb1$ and $lb2$ and $LB2>LB1$.

The above Examples $1$ and $2$ demonstrate that our lower bound, namely, the maximum of $LB1$ and $LB2$, is tighter than the previous lower bounds $lb1$ and $lb2$. They also show that our lower bounds $LB1$ and $LB2$ are complementary in certain cases.

In order to better illustrate our results, we consider the more general case, the random three-qutrit states, $\rho_{ABC}=\sum_{s=1}^{27}\lambda_s|\varphi_s\ra\la\varphi_s|$, where $\lambda_s$ and $|\varphi_s\ra$ denote the $s$th eigenvalue and eigenvector of $\rho_{ABC}$, respectively. In addition, the eigenvalue $\lambda_s$ corresponds to the probability of the system being in the state $|\varphi_s\ra$. And the normalized eigenvector $|\varphi_s\ra$ can constitute an arbitrary unitary operation $\mathcal{V}=\{|\varphi_1\ra,|\varphi_2\ra,\cdots,|\varphi_{27}\ra\}$.
Therefore, we can obtain an arbitrary three-qutrit state by using an arbitrary set of probabilities $\lambda_s$ and an arbitrary unitary operator $\mathcal{V}$.
  Here we adopt an effective approach to generate random states. Firstly, we define a random function $\xi(\theta_1,\theta_2)$, which can generate a independent random real number in a closed interval $[\theta_1,\theta_2]$. The random number $p_s$ is then generated:
\begin{equation*}
p_1=\xi(0,1),~ p_{s+1}=\xi(0,1)p_s,
\end{equation*}
where $s\in\{1,2,\cdots,26\}$. We obtain a set of probabilities in descent order:
\begin{equation*}
\lambda_s=\frac{p_s}{\sum_{s=1}^{27}p_s},~ s=1,\dots,27.
\end{equation*}
Utilizing the random function $\xi(-1,1)$ within a closed interval $[-1,1]$, we randomly generate a twenty-seven-order real matrix $\mathcal{R}$. Based on the real matrix $\mathcal{R}$, a random Hermitian matrix can be obtained as $\mathcal{M}=D+(U^T+U)+\mathrm{i}(L^T+L)$, where $D$, $U$ and $L$, respectively, denote the diagonal, strictly upper and lower triangular parts of the real matrix $\mathcal{R}$, $U^T$ is the transpose of $U$. We obtain twenty-seven normalized eigenvectors $|\varphi_s\ra$ of the Hermitian matrix $\mathcal{M}$ and a random unitary operator $\mathcal{V}$. Therefore, the random three-qutrit states are constructed.

As an illustration, we use 50 sets of three-qutrit states and three-dimensional random measurements to depict the corresponding uncertainty and our bounds in Figure 3. It is seen from Figure 3 that $LB1>lb1$ and $LB2 >lb1$, while $LB1$ and $lb2$ have an uncertain relationship, and $LB2>lb2$. This shows that the maximum of our lower bounds ($LB1$ and $LB2$) outperforms the previous bounds. Additionally, the figure illustrates that $LB1$ and $LB2$ are complementary in certain cases.
\begin{figure}[ht]
  \centering
  \includegraphics[width=8 cm]{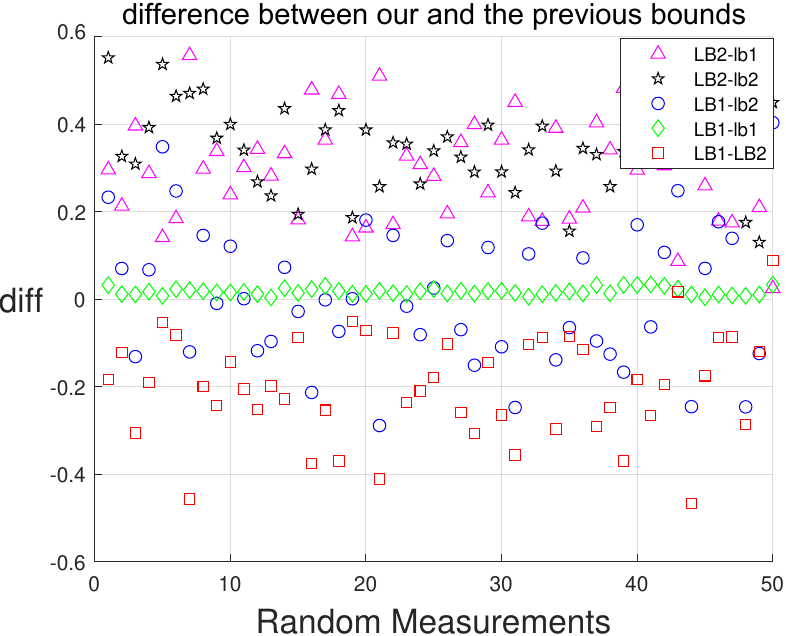}\\
\caption{
The horizontal axis represents the number of random measurements and random states, while the vertical axis represents the difference (diff) between our results and the previous results. The magenta triangle, black star, blue circle, green diamond and the red square represent the values of $LB2-lb1$,$LB2-lb2$,$LB1-lb1$ and $LB1-LB2$, respectively.    \label{fig:Fig3}}
\end{figure}



Therefore, we deduce the following optimal lower bound,
\begin{equation}\label{eq21}
\begin{aligned}
\sum_{t=1}^n \sum_{M_i \in \mathbf{S}_t} S(M_i|B_t) \geq & \frac{1}{m-1} \sum_{i<j} Q^{M_{i,j}} \\
+&  \sum_{t=1}^n \frac{m_t(m_t-1)}{2(m-1)} S(A|B_t) \\
+& \max\{0, \delta_{mn}, \delta_{mn}''\}.
\end{aligned}
\end{equation}

We elucidate the optimal uncertainty relation with one memory scenario ($n=1$, $m_t=m$). (\ref{eq21}) reduces to the case of \( m \) measurements,
\begin{equation*}
\begin{aligned}
\sum_{i=1}^m S\left(M_i \mid B\right) \geq  &\frac{1}{m-1} \sum_{i<j}Q^{M_{i,j}}+\frac{m}{2} S(A|B)\\
+&\max \left\{0, \delta_{m1},\delta_{m1}^{''}\right\},
\end{aligned}
\end{equation*}
where
\begin{equation*}
\begin{aligned}
\delta_{m1}=\frac{m}{2} \mathcal{I}(A: B)-\sum_{i=1}^m \mathcal{I}\left(M_i:B\right)
\end{aligned}
\end{equation*}
and
\begin{equation*}
\begin{aligned}
\delta_{m1}^{''}=&-\frac{1}{m}\omega\beta+(m-1)S(A)-\frac{m}{2}S(A)
+\frac{m}{2}\mathcal{I}(A:B)\\
&-\frac{1}{m-1} \sum_{i<j}Q^{M_{i,j}}-\sum_i^m \mathcal{I}(M_i:B).
\end{aligned}
\end{equation*}
This inequality deftly captures the essence of the uncertainty relations in the single memory scenario, underscoring the intricate interplay between the information metrics and the quantum measurements.

For the case of \( m=n \), \( m \) memories with the cardinality of \(\mathbf{S}_t\) one (\( m_t=1 \)), the inequality (\ref{eq21}) reduces to
\begin{equation*}
\sum_{i}^{m} S(M_i|B_i) \geq \frac{1}{m-1} \sum_{i<j} Q^{M_{i,j}} + \max\{0, \delta_{mm}, \delta_{mm}^{''}\},
\end{equation*}
where
\begin{equation*}
\begin{aligned}
\delta_{mm} = \frac{m}{2} S(A) - \sum_{i}^{m} \mathcal{I}(M_i:B_i)
\end{aligned}
\end{equation*}
and
\begin{equation*}
\begin{aligned}
\delta_{mm}^{''} = &-\frac{1}{m}\omega\beta+(m-1)S(A) - \sum_{i}^{m} \mathcal{I}(M_i:B_i)\\
&- \frac{1}{m-1} \sum_{i<j} Q^{M_{i,j}}.
\end{aligned}
\end{equation*}
This refined inequality encapsulates the uncertainty relations for the case of \( m \) memories, highlighting the intricate balance between the entropic measures and the mutual information terms. The definitions of \(\delta_{mm}\) and \(\delta_{mm}^{''}\) provide a comprehensive framework for understanding the constraints imposed by quantum measurements and the underlying quantum states.

\section{Multipartite QMA-EURs for POVMs}

The most general quantum measurements are given by the positive operator-valued measures (POVMs). A POVM on a system $A$ is a set of positive semidefinite operators $\{E_k\}$ such that $\sum\limits_k E_k = \mathrm{I_A}$ with $\mathrm{I_A}$ the identity operator on the system $A$. Let $X_1 = \{X^1_j\}$ and $X_2 = \{X^2_k\}$ be two arbitrary POVMs on $A$. The following QMA-EURs \cite{coles2014improved},
\begin{equation}\label{eq22}
S(X_1|B) + S(X_2|B) \geq q^{X_{1,2}}(\rho_A) + S(A|B) - f^{X_{1,2}},
\end{equation}
\begin{equation}\label{eq23}
S(X_1|B) + S(X_2|C) \geq q^{X_{1,2}}(\rho_A),
\end{equation}
where $q^{X_{1,2}}(\rho_A) = \mathrm{max}\{q(\rho_A, X_1, X_2), q(\rho_A, X_2, X_1)\}$ with
$q(\rho_A, X_1, X_2) = -\sum\limits_j p^{X_1}_j \log_2 h_j(X_1, X_2)$ and
$q(\rho_A, X_2, X_1) = -\sum\limits_k p^{X_2}_k \log_2 h_k(X_2, X_1)$.
Here, $p^{X_1}_j = \mathrm{tr}(X^1_j \rho_A)$, $p^{X_2}_k = \mathrm{tr}(X^2_k \rho_A)$ and $h_j(X_1, X_2) = \|\sum\limits_k X^2_k X^1_j X^2_k\|_{\infty}$, $h_k(X_2, X_1) = \|\sum\limits_j X^1_j X^2_k X^1_j\|_{\infty}$, where $\|\cdot\|_{\infty}$ denotes the infinity norm, $f^{X_{1,2}}= \mathrm{min}\{S(A|BX_1), S(A|BX_2)\}$, $S(A|BX_1)$ and $S(A|BX_2)$ are the conditional von Neumann entropies of ${\rho}_{X_1AB}$ and ${\rho}_{X_2AB}$, respectively \cite{coles2014improved}. Notably, for orthonormal bases (rank-one projective POVMs), $f^{X_{1,2}} = 0$. In the absence of quantum memory $B$, (\ref{eq22}) reduces to
\begin{equation}\label{eq24}
H(X_1) + H(X_2) \geq q^{X_{1,2}}(\rho_A) + S(A) - \tilde{f}^{X_{1,2}},
\end{equation}
where $\tilde{f}^{X_{1,2}} = \mathrm{min}\{S(A|X_1), S(A|X_2)\}$.

Building on the concept presented in \cite{PhysRevA.108.012211}, we consider a scenario where Alice, Bob$_1$, Bob$_2$, $\dots$, Bob$_n$ agree on a set of generalized measurements (POVMs) $\mathbf{X} = \{X_i\}_{i=1}^m$. We extend the QMA-EURs (\ref{eq22}) and (\ref{eq23}) to encompass $m$ measurements in the context of $n$ memories.

\begin{theorem}\label{qmaeurthm4}
Let $X_1, X_2,\cdots X_m$ be $m$ POVMs. We have the following QMA-EUR for $m$ POVMs within the context of $n$ memories,
\begin{equation}\label{eq25}
\begin{aligned}
&\sum_{t=1}^n\sum_{X_i\in \mathbf{S}_t} S(X_i|B_t)\\
\geq & \frac{1}{m-1}
\sum_{t=1}^{n} \frac{m_t(m_t-1)}{2}S(A|B_t)\\
+&\frac{1}{m-1} \sum_{i<j}q^{X_{i,j}}(\rho_A)
-\frac{1}{m-1}\sum_{t=1}^{n} \mathop{\sum}_{\substack{X_{i'},X_{j'}\in S_t  \\i'<j'}}f^{X_{i',j'}}\\
+&\max\{0,\kappa_{mn}\},
\end{aligned}
\end{equation}
where
\begin{equation*}
\begin{aligned}
\kappa_{mn}=&\frac{m(m-1)-\sum\limits_{t=1}^n m_t(m_t-1)}{2(m-1)}S(A)\\
+&\sum_{t=1}^n\frac{m_t(m_t-1)}{2(m-1)}\mathcal{I}(A:B_t)
-\sum_{t=1}^n\sum_{X_i\in \mathbf{S}_t} \mathcal{I}(X_i:B_t)\\
+&\frac{1}{m-1}\left[\sum_{t=1}^n\mathop{\sum}_{\substack{X_{i'},X_{j'}\in S_t  \\i'<j'}}f^{X_{i',j'}}-\sum_{i<j}^m\tilde{f}^{X_{i,j}}\right],
\end{aligned}
\end{equation*}
with $f^{X_{i',j'}}=\min\{S(A|BX_{i'}),S(A|BX_{j'})\}$ and $\tilde{f}^{X_{i,j}}=\min\{S(A|X_i),S(A|X_j)\}$.
\end{theorem}

\noindent \textit{Proof.} By applying the same method as in the proof of Theorem 2, we obtain
\small{\begin{equation}\label{eq26}
\begin{aligned}
\sum_{t=1}^n\sum_{X_i\in \mathbf{S}_t} S(X_i|B_t)\geq& \frac{1}{m-1}\sum_{i<j}q^{X_{i,j}}(\rho_A)
+\frac{m}{2}S(A)\\
-&\sum_{t=1}^n\sum_{X_i\in \mathbf{S}_t} \mathcal{I}(X_i:B_t)-\frac{1}{m-1}\sum_{i<j}^m\tilde{f}^{X_{i,j}}.
\end{aligned}
\end{equation}}
Combining inequalities (\ref{eq22}) and (\ref{eq23}), we obtain
\begin{equation*}
S(X_i|B_t) + S(X_j|B_s) \geq \left\{
\begin{aligned}
& q^{X_{i,j}}(\rho_A), \quad t \neq s \\
& q^{X_{i,j}}(\rho_A) + S(A|B_t) - f^{X_{i,j}},  t = s.
\end{aligned}
\right.
\end{equation*}
Employing the same techniques as in the proof of Theorem 2, we have,
\begin{equation}\label{eq27}
\begin{aligned}
\sum_{t=1}^n \sum_{X_i \in \mathbf{S}_t} S(X_i|B_t) \geq & \frac{1}{m-1} \sum_{t=1}^{n} \frac{m_t(m_t-1)}{2} S(A|B_t) \\
+& \frac{1}{m-1} \sum_{i<j} q^{X_{i,j}}(\rho_A) \\
-& \frac{1}{m-1} \sum_{t=1}^{n} \mathop{\sum}_{\substack{X_{i'},X_{j'} \in S_t \\ i' < j'}} f^{X_{i',j'}},
\end{aligned}
\end{equation}
where $m_t$ is the cardinality of $\mathbf{S}_t$ and $f^{X_{i',j'}} = \mathrm{min}\{S(A|BX_{i'}), S(A|BX_{j'})\}$.
From (\ref{eq26}), (\ref{eq27}) and that $S(A|B_t)=S(A)-\mathcal{I}(A:B_t)$, we prove the theorem.
$\Box$

In particular, for a bipartite state $\rho_{AB}$ subjected to $m$ measurements on particle $A$ (i.e., $n=1, m_t=m$), Theorem 4 simplifies to:
\begin{equation*}
\begin{aligned}
\sum_{i=1}^m S\left(X_i \mid B\right) \geq  &\frac{1}{m-1}\sum_{i<j}q^{X_{i,j}}(\rho_A)+\frac{m}{2} S(A|B)\\
-&\frac{1}{m-1} \sum_{i<j}f^{X_{i,j}}
+\max \left\{0, \kappa_{m1}\right\},
\end{aligned}
\end{equation*}
where $$\kappa_{m1}=\displaystyle \frac{m}{2} \mathcal{I}(A: B)-\sum_{i=1}^m \mathcal{I}(X_i:B)+\frac{1}{m-1}\sum_{i<j}^m\left[f^{X_{i,j}}-\tilde{f}^{X_{i,j}}\right]$$.

If the cardinality of $\mathbf{S}_t$ is set to be $1$ ($m_t=1$, $n=m$), Theorem 4 further simplifies to
\begin{equation*}
\sum_{i}^m S(X_i|B_i)\geq \frac{1}{m-1}\sum_{i<j}q^{X_{i,j}}(\rho_A)+\max\{0,\kappa_{mm}\},
\end{equation*}
where
$\kappa_{mm}=\displaystyle \frac{m}{2}S(A)-\sum_{i}^m \mathcal{I}(X_i:B_i)-\frac{1}{m-1}\sum\limits_{i<j}\tilde{f}^{X_{i,j}}$.

\section{Applications}
\subsection{Uncertainty relation for quantum coherence}

Coherence is a fundamental issue in quantum physics and a crucial resource for quantum information processing. In 2014 Baumgratz \textit{et al.} \cite{PhysRevLett.113.140401} introduced a framework for quantifying coherence. For a fixed orthonormal basis $\{|i\rangle\}^d_{i=1}$ of the $d$-dimensional Hilbert space $\mathcal{H}$, density matrices that are diagonal in this basis are termed incoherent. We denote the set of incoherent quantum states as $\mathcal{I}=\{\sigma|\sigma=\sum\limits_{i=1}^d\lambda_i|i\rangle\langle i| \}$.

For a bipartite system $AB$, an $A$-incoherent state (also known as an incoherent-quantum state \cite{PhysRevLett.116.070402}) was proposed in \cite{ma2016converting} with respect to the basis (${|i\rangle_A}$) of subsystem $A$ . This is also referred to as unilateral coherence \cite{PhysRevA.95.042328,dolatkhah2019tightening}. We denote the set of $A$-incoherent states as $\mathcal{I}_{B|A}$,
\begin{equation*}
\mathcal{I}_{B|A}=\{\sigma_{AB}|\sigma_{AB}=\sum_{i}p_i|i\rangle\langle i|\otimes\sigma_{B|i}\},
\end{equation*}
where $\sigma_{B|i}$ represents an arbitrary state of the subsystem $B$ \cite{dolatkhah2019tightening}.
Analogous to the definition of the relative entropy coherence \cite{PhysRevLett.113.140401}, the relative entropy of unilateral coherence $C^{B|A}_r$ is given by \cite{dolatkhah2019tightening,PhysRevLett.116.070402},
\begin{equation*}
\begin{aligned}
C^{B|A}_r(\rho_{AB})=& \mathop{\mathrm{min}}\limits_{\sigma_{AB}\in \mathcal{I}_{B|A}} S(\rho_{AB}\|\sigma_{AB})=S(\rho_{AB}\|\Delta_A(\rho_{AB}))\\
=&S(\Delta_A(\rho_{AB}))-S(\rho_{AB}),
\end{aligned}
\end{equation*}
where $\Delta_A(\rho_{AB})=\sum\limits_i(|i\rangle \langle i|\otimes \mathrm{I_{B}})\rho_{AB}(|i\rangle \langle i|\otimes \mathrm{I_{B}})=\sum\limits_i p_i|i\rangle\langle i|\otimes\rho_{B|i}$, $p_i=\mathrm{tr}_{AB}((|i\rangle \langle i|\otimes \mathrm{I_{B}})\rho_{AB}(|i\rangle \langle i|\otimes \mathrm{I_{B}}))$ and $\rho_{B|i}=\mathrm{tr}_{A}((|i\rangle \langle i|\otimes \mathrm{I_{B}})\rho_{AB}(|i\rangle \langle i|\otimes \mathrm{I_{B}}))/p_i$.

For a bipartite state $\rho_{AB}$, measurements $M$ given by orthonormal bases $|q_i\rangle$ on subsystem $A$ give rise to the post-measurement state,
\begin{equation*}
\rho_{MB}=\sum_iq_i|q_i\rangle\langle q_i|\otimes\rho_{B|q_i},
\end{equation*}
$\rho_{MB}$ is an $A$-incoherent state with respect to the ${|q_i\rangle}$ basis. The trade-off relation between measurement uncertainty and the relative entropy of unilateral coherence with respect to the measurement basis ${|q_i\rangle}$ is given by \cite{dolatkhah2019tightening},
\begin{equation}\label{eq28}
C^{M}_r(\rho_{AB})=S(\rho_{AB}\|\rho_{MB})=S(M|B)-S(A|B).
\end{equation}

Consider incompatible quantum measurements $M_i$ $(i=1,2,\cdots,m)$ on subsystem $A$ in the context of $n$ memories, we obtain
\begin{equation}\label{eq29}
\sum_{t=1}^n\sum_{M_i\in \mathbf{S}_t}C^{M_i}_r(\rho_{AB_t})=\sum_{t=1}^n\sum_{M_i\in \mathbf{S}_t}S(M_i|B_t)-\sum_{t=1}^nm_tS(A|B_t).
\end{equation}
From (\ref{eq29}) with (\ref{eq21}) we derive a tight uncertainty relation based on coherence,
\begin{equation}\label{eq30}
\begin{aligned}
\sum_{t=1}^n\sum_{M_i\in \mathbf{S}_t}C^{M_i}_r(\rho_{AB_t})\geq&
\frac{1}{m-1} \sum_{i<j}Q^{M_{i,j}}\\
+&\sum_{t=1}^{n} \frac{m_t(m_t-2m+1)}{2(m-1)}S(A|B_t)\\
+&\max\{0,\delta_{mn},\delta_{mn}^{''}\}.
\end{aligned}
\end{equation}

We employ an analogous approach to generate a random two-qutrit state  $\rho_{AB}$. Unlike the three-qutrit case, $\rho_{AB}$ has nine eigenvalues and eigenvectors here. The procedure begins with generating a random $9\times9$ real matrix $R_1$, from which we construct the Hermitian matrix
$\mathcal{M}_1=D_1+(U_1^T+U_1)+\mathrm{i}(L_1^T+L_1)$. Here $D_1$ contains the diagonal elements of $\mathcal{R}_1$, while $U_1$ and $L_1$ represent its strictly upper and lower triangular components, respectively. Diagonalizing $\mathcal{M}_1$ yields nine normalized eigenvectors, which are then used to form a random unitary matrix $\mathcal{V}_1$. The remaining steps mirror the three-qutrit construction method, ultimately producing the desired two-qutrit state.

 \textbf{\textit{Example 3}} Consider $10^4$ two-qutrit random states $\rho_{AB}$ and three measurements $M_1$, $M_2$ and $M_3$ with the corresponding  eigenvectors
$|\varphi^1_1 \rangle=(\frac{\sqrt{2}}{2},\frac{\sqrt{2}}{2},0)^T$, $|\varphi^1_2 \rangle=(-\frac{\sqrt{2}}{2},\frac{\sqrt{2}}{2},0)^T$, $|\varphi^1_3 \rangle=(0,0,1)^T$,
$|\varphi^2_1 \rangle=(\frac{\sqrt{6}}{6},-\frac{\sqrt{6}}{6},\frac{\sqrt{6}}{3})^T$, $|\varphi^2_2 \rangle=(-\frac{\sqrt{3}}{3},\frac{\sqrt{3}}{3},\frac{\sqrt{3}}{3})^T$, $|\varphi^2_3 \rangle=(\frac{\sqrt{2}}{2},\frac{\sqrt{2}}{2},0)^T$,
$|\varphi^3_1 \rangle=(\frac{\sqrt{2}}{2},0,\frac{\sqrt{2}}{2})^T$, $|\varphi^3_2 \rangle=(0,1,0)^T$, $|\varphi^3_3 \rangle=(-\frac{\sqrt{2}}{2},0,\frac{\sqrt{2}}{2})^T$.
In this scenario ($m_t=m=3$, $n=1$), our inequality (\ref{eq30}) gives rise to
\begin{equation}
\begin{aligned}\label{eq31}
C^{M_1}_r(\rho_{AB})+C^{M_2}_r(\rho_{AB})+C^{M_3}_r(\rho_{AB})
&\geq \frac{3}{2}-\frac{3}{2}S(A|B)\\
&+\max\{0,\delta_{31},\delta_{31}^{''}\},
\end{aligned}
\end{equation}
with
\begin{equation*}
\begin{aligned}
\delta_{31}=\frac{3}{2} \mathcal{I}(A: B)-\sum_{i=1}^3 \mathcal{I}\left(M_i:B\right),
\end{aligned}
\end{equation*}
\begin{equation*}
\begin{aligned}
\delta_{31}^{''}=&-\frac{1}{3}\omega\beta+\frac{1}{2}S(A)
+\frac{3}{2}\mathcal{I}(A:B)-\frac{1}{2} \sum_{i<j}Q^{M_{i,j}}\\
&-\sum_i^3 \mathcal{I}(M_i:B).
\end{aligned}
\end{equation*}
\begin{figure}[tp]
  \centering
  \includegraphics[width=5 cm]{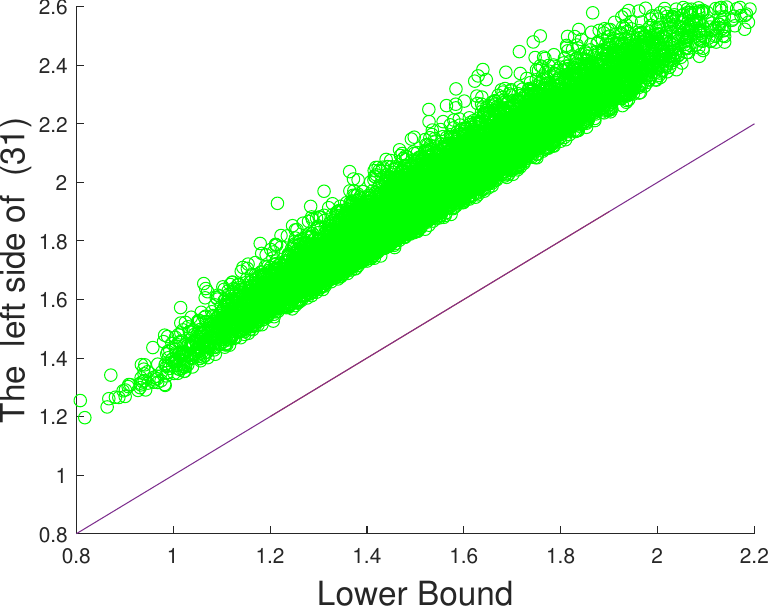}\\
 \caption{ Uncertainty relation for quantum coherence under three multiple measurements for $10^4$ randomly generated two-qutrit states. The horizontal axis represents the lower bound in inequality (\ref{eq31}), while the vertical axis represents the value of  $C^{M_1}_r(\rho_{AB})+C^{M_2}_r(\rho_{AB})+C^{M_3}_r(\rho_{AB})$.  The red solid line is the proportional function with a slope of unity, displaying that the lower bound in inequality (\ref{eq31}) is equal to $C^{M_1}_r(\rho_{AB})+C^{M_2}_r(\rho_{AB})+C^{M_3}_r(\rho_{AB})$.
The green circles denotes the value of  $C^{M_1}_r(\rho_{AB})+C^{M_2}_r(\rho_{AB})+C^{M_3}_r(\rho_{AB})$.}
\end{figure}
In Figure 4, the coherence-based uncertainty relation (\ref{eq31}) is illustrated for $10^4$ randomly generated two-qutrit states and three measurements.

\subsection{Quantum secret key rate}
The entropy uncertainty relation is tightly related to the security of quantum key distribution (QKD): a tighter entropy uncertainty relation ensures a higher quantum secret key rate (QSKR), thereby enhancing the security of QKD. Consider two honest participants, Alice and Bob, who communicate over a public channel to share a secret key which is kept confidential from any third-party eavesdropper, Charles. In 2005, Devetak and Winter \cite{doi:10.1098/rspa.2004.1372} demonstrated that the QSKR \( K \) that Alice and Bob can extract is lower bounded by
\begin{equation}\label{eq32}
K \geq S(M_2|C) + S(M_2|B).
\end{equation}

The eavesdropper (Charles) prepares a quantum state \(\rho_{ABC}\) and distributes subsystems \(A\) and \(B\) to Alice and Bob, respectively, while retains \(C\). Combining inequality (\ref{eq32}) with inequality (\ref{eq6}), we obtain \( K \geq Q^{M_{1,2}} - S(M_1|B) - S(M_2|B) \). Furthermore, since measurements cannot decrease the system's entropy \cite{berta2010uncertainty,ming2020improved, wu2022tighter}, i.e.,
\begin{equation}\label{eq33}
S(M_1|B) \leq S(M_1|M^{'}_1),~~~ S(M_2|B) \leq S(M_2|M^{'}_2),
\end{equation}
we derive
\begin{equation}\label{eq34}
K \geq Q^{M_{1,2}} - S(M_1|M^{'}_1) - S(M_2|M^{'}_2),
\end{equation}
where $M_1$ and $M_2$ ($M^{'}_1$ and $M^{'}_2$ ) are the measurements chose by Alice (Bob), $S(M_1|B)$ is the same as in inequality (\ref{eq3}) and $S(M_1|M^{'}_1)=S(\rho_{M_1 M^{'}_1}) -S(\rho_{M^{'}_1})$ (similarly for $S(M_2|M^{'}_2)$) with $\rho_{M_1 M^{'}_1}=\sum_{j'}(\mathrm{I_{A}}\otimes |\varphi_{j'}^1\rangle\langle \varphi_{j'}^1|)\rho_{M_{1}B}(\mathrm{I_{A}}\otimes |\varphi_{j'}^1\rangle\langle \varphi_{j'}^1|)$, $\rho_{M^{'}_1}=\mathrm{tr}_{M_1}\rho_{M_1 M^{'}_1}$.

Taking into account the lower bound from Theorem 1, a new lower bound on generating the QSKR can be established,
\begin{equation}\label{eq35}
\widetilde{K} \geq Q^{M_{1,2}} - S(M_1|B) - S(M_2|B) + \max\{0, \Delta\}.
\end{equation}
Moreover, utilizing the inequalities (\ref{eq33}), the lower bound of the QSKR can be refined to
\begin{equation}\label{eq36}
\widetilde{K} \geq Q^{M_{1,2}} - S(M_1|M^{'}_1) - S(M_2|M^{'}_2) + \max\{0, \Delta\}.
\end{equation}
In this context, the lower bound of inequality (\ref{eq36}) is more stringent than that of inequality (\ref{eq34}) due to \(\max\{0, \Delta\} \geq 0\). Consequently, it is validated that \(\widetilde{K} \geq K\), confirming that the results indeed bolster the security of quantum key distribution protocols.

\section{Conclusion}
We have proposed a tripartite QMA-EUR and generalized it to the case of multiple measurements within multipartite systems by using two distinct approaches. An optimal lower bound has been derived. Additionally, explicit examples have been presented, demonstrating that our lower bounds are tighter than those in \cite{PhysRevA.108.012211}. Furthermore, we have extended our results to encompass arbitrary positive-operator-valued measures (POVMs). Regarding the QMA-EURs, we have also derived uncertainty relations for the sum of relative entropy of unilateral coherence. Notably, our tripartite QMA-EUR indeed enhances the quantum key secret rates, thereby strengthening the security of the quantum key distribution protocol. Our findings may illuminate new avenues for the optimal entropic uncertainty relation and its applications.

\bigskip
{\bf Acknowledgements}
This work was supported by National Natural Science Foundation of China (Grant Nos. 12075159, 12171044); Natural Science Foundation of Hunan Province (Grant No. 2025JJ60025); the specific research fund of the Innovation Platform for Academicians of Hainan Province and Changsha University of Science and Technology (Grant No. 097000303923); the Disciplinary funding of Beijing Technology and Business University; Scientific Research Project of the Education Department of Hunan Province (Grant No.24B0298).
\nocite{*}
\bibliography{2025-pra-XC-Li}

\end{document}